\pdfoutput=1
\documentclass[12pt,a4paper]{article}

\usepackage{lscape}
\usepackage{ifthen} 
\newboolean{pdflatex}
\setboolean{pdflatex}{true} 
\usepackage{url}
\newboolean{articletitles}
\setboolean{articletitles}{true} 

\newboolean{uprightparticles}
\setboolean{uprightparticles}{false} 

\usepackage{multirow}
\usepackage{verbatim}

\newcommand{\BsDtphi}{\Bs\rightarrow{{\ensuremath{\Dbar{}^{(*)0}}}\xspace\phi}}
\newcommand{\Dtphi}{{\ensuremath{\Dbar{}^{(*)0}}}\xspace\phi}
\newcommand{\BzDtphi}{\Bz\rightarrow{{\ensuremath{\Dbar{}^{(*)0}}}\xspace\phi}}
\newcommand{\BzDtomega}{\Bz\rightarrow {{\ensuremath{\Dbar{}^{(*)0}}}\xspace\omega}}

\newcommand{\BzDbkk}{\Bz\rightarrow \Dzb\Kp\Km}

\def\paperauthors{LHCb collaboration} 
\def\paperasciititle{Bs to D(star) phi paper draft} 
\def\papertitle{Evidence for the decays $\BzDtphi$ and updated measurements of the branching fractions of the $\BsDtphi$ decays} 
\def\paperkeywords{{High Energy Physics}, {LHCb}} 
\def\papercopyright{\the\year\ CERN for the benefit of the LHCb collaboration} 
\def\paperlicence{CC BY 4.0 licence}
\def\paperlicenceurl{https://creativecommons.org/licenses/by/4.0/}

\usepackage[top=1in, bottom=1.25in, left=1in, right=1in]{geometry}

\usepackage{microtype}
\usepackage{lineno}  
\usepackage{xspace} 
\usepackage{caption} 

\usepackage{graphicx}  
\usepackage{color}
\usepackage{colortbl}
\graphicspath{{./figs/}} 

\usepackage{amsmath} 
\usepackage{amssymb}
\usepackage{amsfonts}
\usepackage{upgreek} 

\newcommand*\patchAmsMathEnvironmentForLineno[1]{%
\expandafter\let\csname old#1\expandafter\endcsname\csname #1\endcsname
\expandafter\let\csname oldend#1\expandafter\endcsname\csname
end#1\endcsname
 \renewenvironment{#1}%
   {\linenomath\csname old#1\endcsname}%
   {\csname oldend#1\endcsname\endlinenomath}%
}
\newcommand*\patchBothAmsMathEnvironmentsForLineno[1]{%
  \patchAmsMathEnvironmentForLineno{#1}%
  \patchAmsMathEnvironmentForLineno{#1*}%
}
\AtBeginDocument{%
\patchBothAmsMathEnvironmentsForLineno{equation}%
\patchBothAmsMathEnvironmentsForLineno{align}%
\patchBothAmsMathEnvironmentsForLineno{flalign}%
\patchBothAmsMathEnvironmentsForLineno{alignat}%
\patchBothAmsMathEnvironmentsForLineno{gather}%
\patchBothAmsMathEnvironmentsForLineno{multline}%
\patchBothAmsMathEnvironmentsForLineno{eqnarray}%
}

\usepackage{hyperxmp}

\usepackage[pdftex,
            pdfauthor={\paperauthors},
            pdftitle={\paperasciititle},
            pdfkeywords={\paperkeywords},
            pdfcopyright={Copyright (C) \papercopyright},
            pdflicenseurl={\paperlicenceurl}]{hyperref}

\usepackage[colorinlistoftodos,textsize=scriptsize]{todonotes}

\usepackage[bottom,flushmargin,hang,multiple]{footmisc}

\usepackage[all]{hypcap} 

\usepackage{xspace} 
\usepackage{upgreek}

\def\lhcb   {\mbox{LHCb}\xspace}

\def\MagUp {\mbox{\em Mag\kern -0.05em Up}\xspace}

\ifthenelse{\boolean{uprightparticles}}%
{

 \def\Ppi         {\ensuremath{\uppi}\xspace}

 \def\Ppsi        {\ensuremath{\uppsi}\xspace}

 \def\PDelta      {\ensuremath{\Delta}\xspace}                 
 \def\PXi         {\ensuremath{\Xi}\xspace}                 
 \def\PLambda     {\ensuremath{\Lambda}\xspace}                 
 \def\PSigma      {\ensuremath{\Sigma}\xspace}                 
 \def\POmega      {\ensuremath{\Omega}\xspace}                 
 \def\PUpsilon    {\ensuremath{\Upsilon}\xspace}
 \let\oldPi\Pi
 \def\PPi         {\ensuremath{\oldPi}\xspace}

 \def\PB      {\ensuremath{\mathrm{B}}\xspace}                 
                  
 \def\PD      {\ensuremath{\mathrm{D}}\xspace}

 \def\PJ      {\ensuremath{\mathrm{J}}\xspace}                 
 \def\PK      {\ensuremath{\mathrm{K}}\xspace}

 \def\PW      {\ensuremath{\mathrm{W}}\xspace}

 \def\Pb      {\ensuremath{\mathrm{b}}\xspace}                 
 \def\Pc      {\ensuremath{\mathrm{c}}\xspace}

 \def\Pi      {\ensuremath{\mathrm{i}}\xspace}

 \def\Pp      {\ensuremath{\mathrm{p}}\xspace}

 \def\Ps      {\ensuremath{\mathrm{s}}\xspace}

 \def\thebaroffset{0.0em}
}
{

 \def\Ppi         {\ensuremath{\pi}\xspace}

 \def\Ppsi        {\ensuremath{\psi}\xspace}                 
                  
 \mathchardef\PDelta="7101
 \mathchardef\PXi="7104
 \mathchardef\PLambda="7103
 \mathchardef\PSigma="7106
 \mathchardef\POmega="710A
 \mathchardef\PUpsilon="7107
 \mathchardef\PPi="7105
                  
 \def\PB      {\ensuremath{B}\xspace}                 
                  
 \def\PD      {\ensuremath{D}\xspace}

 \def\PJ      {\ensuremath{J}\xspace}                 
 \def\PK      {\ensuremath{K}\xspace}

 \def\PW      {\ensuremath{W}\xspace}

 \def\Pb      {\ensuremath{b}\xspace}                 
 \def\Pc      {\ensuremath{c}\xspace}

 \def\Pi      {\ensuremath{i}\xspace}

 \def\Pp      {\ensuremath{p}\xspace}

 \def\Ps      {\ensuremath{s}\xspace}

 \def\thebaroffset{0.18em}
}
\newcommand{\offsetoverline}[2][\thebaroffset]{\kern #1\overline{\kern -#1 #2}}%

\makeatletter
\ifcase \@ptsize \relax
  \newcommand{\miniscule}{\@setfontsize\miniscule{4}{5}}
\or
  \newcommand{\miniscule}{\@setfontsize\miniscule{5}{6}}
\or
  \newcommand{\miniscule}{\@setfontsize\miniscule{5}{6}}
\fi
\makeatother

\DeclareRobustCommand{\optbar}[1]{\shortstack{{\miniscule (\rule[.5ex]{1.25em}{.18mm})}
  \\ [-.7ex] $#1$}}

\def\squark    {{\ensuremath{\Ps}}\xspace}

\def\cquark    {{\ensuremath{\Pc}}\xspace}

\def\bquark    {{\ensuremath{\Pb}}\xspace}

\def\pion   {{\ensuremath{\Ppi}}\xspace}
\def\piz    {{\ensuremath{\pion^0}}\xspace}
\def\pip    {{\ensuremath{\pion^+}}\xspace}
\def\pim    {{\ensuremath{\pion^-}}\xspace}
\def\pipm   {{\ensuremath{\pion^\pm}}\xspace}
\def\pimp   {{\ensuremath{\pion^\mp}}\xspace}

\def\kaon    {{\ensuremath{\PK}}\xspace}

\def\KorKbar {\kern \thebaroffset\optbar{\kern -\thebaroffset \PK}{}\xspace}

\def\Kp      {{\ensuremath{\kaon^+}}\xspace}
\def\Km      {{\ensuremath{\kaon^-}}\xspace}
\def\Kpm     {{\ensuremath{\kaon^\pm}}\xspace}

\def\Kstarz  {{\ensuremath{\kaon^{*0}}}\xspace}

\def\Dbar    {{\ensuremath{\offsetoverline{\PD}}}\xspace}
\def\D       {{\ensuremath{\PD}}\xspace}

\def\DorDbar {\kern \thebaroffset\optbar{\kern -\thebaroffset \PD}\xspace}
\def\Dz      {{\ensuremath{\D^0}}\xspace}
\def\Dzb     {{\ensuremath{\Dbar{}^0}}\xspace}
\def\Dp      {{\ensuremath{\D^+}}\xspace}
\def\Dm      {{\ensuremath{\D^-}}\xspace}

\def\DpDm    {\ensuremath{\Dp {\kern -0.16em \Dm}}\xspace}
\def\Dstar   {{\ensuremath{\D^*}}\xspace}

\def\Dstarzb {{\ensuremath{\Dbar{}^{*0}}}\xspace}

\def\theDstarm{{\ensuremath{\D^{*}(2010)^{-}}}\xspace}

\def\Ds      {{\ensuremath{\D^+_\squark}}\xspace}

\def\Dsm     {{\ensuremath{\D^-_\squark}}\xspace}

\def\Dsmp    {{\ensuremath{\D^{\mp}_\squark}}\xspace}

\def\B       {{\ensuremath{\PB}}\xspace}
\def\Bbar    {{\ensuremath{\offsetoverline{\PB}}}\xspace}

\def\BorBbar {\kern \thebaroffset\optbar{\kern -\thebaroffset \PB}\xspace}
\def\Bz      {{\ensuremath{\B^0}}\xspace}

\def\Bd      {{\ensuremath{\B^0}}\xspace}

\def\BdorBdbar {\kern \thebaroffset\optbar{\kern -\thebaroffset \Bd}\xspace}

\def\Bpm     {{\ensuremath{\B^\pm}}\xspace}

\def\Bs      {{\ensuremath{\B^0_\squark}}\xspace}
\def\Bsb     {{\ensuremath{\Bbar{}^0_\squark}}\xspace}
\def\BsorBsbar {\kern \thebaroffset\optbar{\kern -\thebaroffset \Bs}\xspace}

\def\jpsi     {{\ensuremath{{\PJ\mskip -3mu/\mskip -2mu\Ppsi}}}\xspace}

\def\Y#1S{\ensuremath{\PUpsilon{(#1S)}}\xspace}

\def\proton      {{\ensuremath{\Pp}}\xspace}

\def\Lz          {{\ensuremath{\PLambda}}\xspace}

\def\LorLbar     {\kern \thebaroffset\optbar{\kern -\thebaroffset \PLambda}\xspace}

\def\Xires       {{\ensuremath{\PXi}}\xspace}

\def\Lb           {{\ensuremath{\Lz^0_\bquark}}\xspace}

\def\Xib          {{\ensuremath{\Xires_\bquark}}\xspace}

\def\BF         {{\ensuremath{\mathcal{B}}}\xspace}

\newcommand{\decay}[2]{\ensuremath{#1\!\to #2}\xspace} 

\def\to                 {\ensuremath{\rightarrow}\xspace}

\def\CP                {{\ensuremath{C\!P}}\xspace}

\def\AT#1     {\ensuremath{A_{\mathrm{T}}^{#1}}\xspace}           

\def\C#1      {\ensuremath{\mathcal{C}_{#1}}\xspace}                       
\def\Cp#1     {\ensuremath{\mathcal{C}_{#1}^{'}}\xspace}                    
\def\Ceff#1   {\ensuremath{\mathcal{C}_{#1}^{\mathrm{(eff)}}}\xspace}        
\def\Cpeff#1  {\ensuremath{\mathcal{C}_{#1}^{'\mathrm{(eff)}}}\xspace}       
\def\Ope#1    {\ensuremath{\mathcal{O}_{#1}}\xspace}                       
\def\Opep#1   {\ensuremath{\mathcal{O}_{#1}^{'}}\xspace}                    

\newcommand{\nospaceunit}[1]{\ensuremath{\text{#1}}}       
\newcommand{\aunit}[1]{\ensuremath{\text{\,#1}}}       

\newcommand{\tev}{\aunit{Te\kern -0.1em V}\xspace}
\newcommand{\gev}{\aunit{Ge\kern -0.1em V}\xspace}
\newcommand{\mev}{\aunit{Me\kern -0.1em V}\xspace}
\newcommand{\kev}{\aunit{ke\kern -0.1em V}\xspace}
\newcommand{\ev}{\aunit{e\kern -0.1em V}\xspace}
 
\newcommand{\mevc}{\ensuremath{\aunit{Me\kern -0.1em V\!/}c}\xspace}
\newcommand{\gevc}{\ensuremath{\aunit{Ge\kern -0.1em V\!/}c}\xspace}
\newcommand{\mevcc}{\ensuremath{\aunit{Me\kern -0.1em V\!/}c^2}\xspace}
\newcommand{\gevcc}{\ensuremath{\aunit{Ge\kern -0.1em V\!/}c^2}\xspace}

\def\mum  {\ensuremath{\,\upmu\nospaceunit{m}}\xspace}

\def\fb   {\ensuremath{\aunit{fb}}\xspace}
\def\invfb   {\ensuremath{\fb^{-1}}\xspace}

\newcommand{\chisq}{\ensuremath{\chi^2}\xspace}

\newcommand{\chisqip}{\ensuremath{\chi^2_{\text{IP}}}\xspace}

\def\gsim{{~\raise.15em\hbox{$>$}\kern-.85em
          \lower.35em\hbox{$\sim$}~}\xspace}
\def\lsim{{~\raise.15em\hbox{$<$}\kern-.85em
          \lower.35em\hbox{$\sim$}~}\xspace}

\def\sqs   {\ensuremath{\protect\sqrt{s}}\xspace}

\def\pt         {\ensuremath{p_{\mathrm{T}}}\xspace}

\def\ptot       {\ensuremath{p}\xspace}

\def\evtgen     {\mbox{\textsc{EvtGen}}\xspace}

\def\geant      {\mbox{\textsc{Geant4}}\xspace}

\def\photos     {\mbox{\textsc{Photos}}\xspace}

\def\pythia     {\mbox{\textsc{Pythia}}\xspace}

\def\tell1  {TELL1\xspace}
\def\ukl1   {UKL1\xspace}

\newcommand{\ie}{\mbox{\itshape i.e.}\xspace}

\newcommand{\lhcborcid}[1]{\href{https://orcid.org/#1}{\hspace*{0.1em}\raisebox{-0.45ex}{\includegraphics[width=1em]{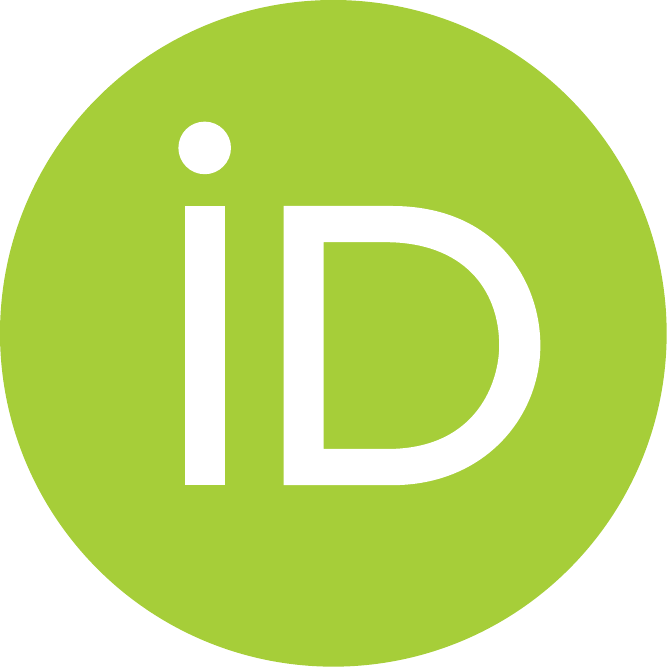}}}}

\usepackage{cite} 
\usepackage{mciteplus}

\usepackage{longtable} 

\begin{document}

\renewcommand{\thefootnote}{\fnsymbol{footnote}}
\setcounter{footnote}{1}


\begin{titlepage}
\pagenumbering{roman}

\vspace*{-1.5cm}
\centerline{\large EUROPEAN ORGANIZATION FOR NUCLEAR RESEARCH (CERN)}
\vspace*{1.5cm}
\noindent
\begin{tabular*}{\linewidth}{lc@{\extracolsep{\fill}}r@{\extracolsep{0pt}}}
\ifthenelse{\boolean{pdflatex}}
{\vspace*{-1.5cm}\mbox{\!\!\!\includegraphics[width=.14\textwidth]{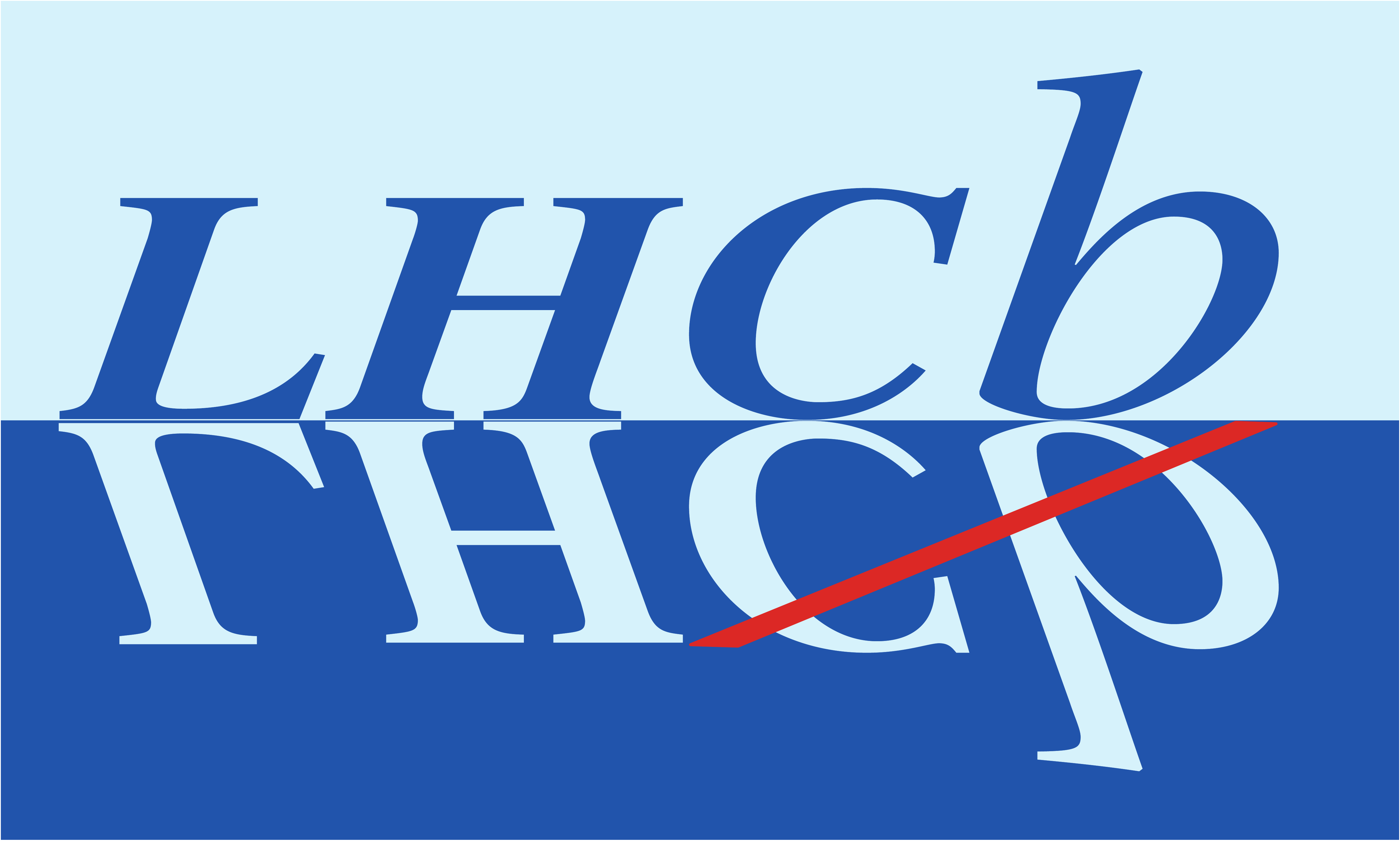}} & &}%
{\vspace*{-1.2cm}\mbox{\!\!\!\includegraphics[width=.12\textwidth]{figs/lhcb-logo.eps}} & &}%
\\
 & & CERN-EP-2023-077 \\  
 & & LHCb-PAPER-2023-003 \\  
 & & \today \\ 
 & & \\
\end{tabular*}

\vspace*{0.5cm}

{\normalfont\bfseries\boldmath\huge
\begin{center}
  \papertitle  
\end{center}
}

\vspace*{2.0cm}

\vspace{-1.5cm}
\begin{center}
\paperauthors\footnote{Authors are listed at the end of this paper.}
\end{center}

\vspace{\fill}
\vspace{-0.5cm}
\begin{abstract}
  \noindent
  
 Evidence for the decays $\Bz\to\Dzb\phi$ and $\Bz\to\Dstarzb\phi$ is reported with a significance of 3.6$\,\sigma$ and 4.3$\,\sigma$, respectively. 
 The analysis employs $pp$ collision data at centre-of-mass energies $\sqs=7$, 8 and 13~TeV collected by the \lhcb detector and corresponding to an integrated luminosity of 9$\invfb$. 
 The branching fractions are measured 
 to be
 \begin{align*}
 \BF(\Bz\to\Dzb\phi)=(7.7\pm2.1\pm0.7\pm0.7)\times10^{-7},\\ 
 \BF(\Bz\to\Dstarzb\phi)=(2.2\pm0.5\pm0.2\pm0.2)\times10^{-6}.
 \end{align*}
In these results, the first uncertainty is statistical, the second systematic, and the third is related to the branching fraction of the $\BzDbkk$ decay, used for normalisation. 
 By combining the branching fractions of the decays $\BzDtphi$ and $\BzDtomega$, the $\omega$-$\phi$ mixing angle $\delta$ is constrained to be \mbox{$\tan^2\delta = (3.6\pm0.7\pm0.4)\times10^{-3}$}, where the first uncertainty is statistical and the second systematic. 
  An updated measurement of the branching fractions of the $\BsDtphi$ decays, 
 which can be used to determine the CKM angle $\gamma$,
 leads to 
 \begin{align*}
 \BF(\Bs\to\Dzb\phi)=(2.30\pm0.10\pm0.11\pm0.20)\times10^{-5},\\
 \BF(\Bs\to\Dstarzb\phi)=(3.17\pm0.16\pm0.17\pm0.27)\times10^{-5}.
  \end{align*}
\end{abstract}

\vspace*{1.0cm}

\begin{center}
 Published in JHEP 10 (2023) 123
\end{center}

\vspace{\fill}

{\footnotesize 
\centerline{\copyright~\papercopyright. \href{\paperlicenceurl}{\paperlicence}.}}
\vspace*{2mm}

\end{titlepage}


\newpage
\setcounter{page}{2}
\mbox{~}

\renewcommand{\thefootnote}{\arabic{footnote}}
\setcounter{footnote}{0}

\cleardoublepage


\pagestyle{plain} 
\setcounter{page}{1}
\pagenumbering{arabic}

\section{Introduction}
\label{sec:Introduction}






Weak nonleptonic decays of $B$ mesons provide a valuable source of information about the relative importance of various flavour-topology amplitudes and rescattering contributions. 
The decay $\BzDtphi$ proceeds at leading order through \PW-exchange diagrams~\cite{Gronau:2008kk}, which are suppressed by the Okubo-Zweig-Iizuka rule~(OZI-suppression)~\cite{Okubo:1963fa,Zweig:352337,Iizuka:1966fk,PhysRevD.64.094507}. The Feynman diagram for the decay can be found in Fig.~\ref{fig: Feyman}. Such OZI-suppressed modes have not been observed yet in $b$-hadron decays~\cite{LHCb-PAPER-2020-033} but were recently observed in charmonium resonance decays $\chi_{cJ}\to\omega\phi$~\cite{BESIII:2018pbx}.
In addition, in the quark model, isoscalar states with the same quantum numbers mix. Strange-nonstrange mixing is expected to be small for isoscalar vector states since they belong to a homochiral multiplet~\cite{Giacosa:2017pos}. As such, the $\omega(782)$ meson is mostly nonstrange and the $\phi(1020)$ meson is mostly strange. The departure from ideal mixing~\cite{PDG2022} and its nature is of particular theoretical interest~\cite{Kucukarslan:2006wk, Qian:2008px, Benayoun:2009im, Volkov:2020jor}.  Significant enhancement by rescattering beyond the $\omega$-$\phi$ mixing through intermediate states is unlikely~\cite{Gronau:2008kk}.
The authors in Ref.~\cite{Gronau:2008kk} predict the branching fraction to be $(2.1\pm0.3)\times10^{-6}$ for $\Bz\to\Dzb\phi$ and $(1.8\pm0.5)\times10^{-6}$ for $\Bz\to\Dstarzb\phi$, which is based on the universal value of $-4.64^{\circ}$ of the $\omega$-$\phi$ mixing angle and the measured OZI-allowed branching fractions.

\begin{figure}[h]
  \begin{center}
    \includegraphics[width=0.49\textwidth]{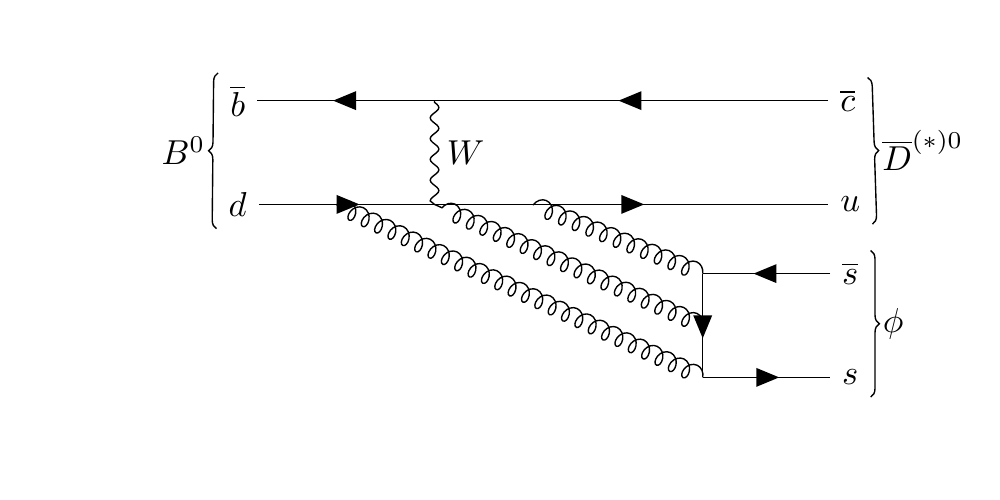}
  \end{center}
  \caption{ Feynman diagram for the decay $\BzDtphi$.
   } 
  \label{fig:Feyman}
\end{figure}

A search for the OZI-suppressed modes $\BzDtphi$ allows the $\omega$-$\phi$ mixing angle to be constrained when combining with measurements of $\BzDtomega$ modes.
A previous analysis~\cite{LHCb-PAPER-2018-015} was published in 2018 by the \lhcb collaboration using a data set corresponding to an integrated luminosity of 3.2~$\invfb$.
An upper limit on the branching fraction of $\Bz\to\Dzb\phi$ was determined to be $\BF(\Bz\to\Dzb\phi)<1.98\times10^{-6}$ at 90\% CL and no measurement for the $\Bz\to\Dstarzb\phi$ mode was provided. A constraint on the $\omega$-$\phi$ mixing angle $\delta$ was set at $|\delta| < 5.2^\circ$ at 90\% CL.

In flavour physics, one of the main goals is precision measurements of the Cabibbo-Kobayashi-Maskawa (CKM)~\cite{Kobayashi:1973fv} angle $\gamma$ in various $B$-meson decay modes. Determinations of $\gamma$ usually exploit the interference of decays that proceed via the $b\to c\bar{u}s$ and $b\to u \bar{c}s$ tree-level amplitudes, in which the determination of the relative weak phase $\gamma$ is not affected by theoretical uncertainties.
At \lhcb, the best precision on the angle $\gamma$ is obtained by combining measurements of many decay modes, which yields $\gamma=(63.8^{+3.5}_{-3.7})^{\circ}$~\cite{LHCb:2022Comb}. This determination dominates the world average of $\gamma$ from tree-level decays. Meanwhile, the measurement of $\gamma$ with $\Bs$ mesons is presently based only on the decay modes $\Bs\to\D^{\mp}_sK^{\pm}$~\cite{LHCB-PAPER-2017-047} and $\Bs\to\D^{\mp}_sK^{\pm}\pip\pim$~\cite{LHCb-PAPER-2020-030}, which leads to a combined constraint $\gamma=(79^{+21}_{-24})^{\circ}$~\cite{LHCb:2022Comb}. Additional methods employing other $\Bs$ decay modes are therefore of interest to improve the precision of the constraint provided by $\Bs$ modes and to verify compatibility with the $B^+$ modes, which dominate the precision of $\gamma$.
The decays $\BsDtphi$ were previously observed by the \lhcb collaboration in 2013~\cite{LHCb-PAPER-2013-035}, and in 2018~\cite{LHCb-PAPER-2018-015}, with data samples corresponding to integrated luminosities of 1~$\invfb$ and 3~$\invfb$, respectively. Theoretical predictions for their branching fractions are given in Refs.~\cite{Li:2008ts,Zhou:2015jba,Talebtash:2016tym,Kaur:2017naw,Li:2020zng,Ahmed:2021oft}. The potential sensitivity of these decays to $\gamma$ was studied in Ref.~\cite{Ao:2020cwh}, which showed that a precision of about $8^{\circ}$ to $19^{\circ}$ is achievable, by using the full dataset collected by the \lhcb collaboration.

In this paper, searches for $\BzDtphi$ decays and improved measurements of the branching fractions of the  $\BsDtphi$  modes are presented.\footnote{Charge conjugation is implied throughout this paper.} Comparing with the previous work, in addition to the increase of the sample size, selection criteria are optimised in the $\phi$ region and more signal candidates are obtained with similar purity. The $\phi$ mesons are reconstructed through decays to $\Kp\Km$ mesons and the $\Dzb$ mesons through decays to $\Kp\pim$ mesons. The $B^0_{(s)}\to\Dstarzb\phi$ decay is partially reconstructed without consideration of the neutral $\piz$ or photon from the $\Dstarzb$ decay, thus allowing a large detection efficiency. The branching fractions are determined relative to that of the decay mode $\Bz\to\Dzb\Kp\Km$~\cite{LHCb-PAPER-2018-014,LHCb-PAPER-2018-015}, which is used as a normalisation channel and allows the cancellation of most of the systematic uncertainties related to the detection efficiency. A measurement of the $\omega$-$\phi$ mixing angle is also performed.

The analysis is based on proton-proton ($pp$) collision data recorded with the LHCb detector between 2011 and 2018 at centre-of-mass energies of 7, 8 and 13~\tev, corresponding to an integrated luminosity of about 9$\invfb$. To consider different efficiencies and cross-sections at various centre-of-mass energies, the dataset is split into two samples: the 2011-2012 data sample (denoted Run 1, about 3\invfb), with collision centre-of-mass energies of $\sqrt{s}=7\tev$~(2011) and $\sqrt{s}=8\tev$~(2012); and the 2015-2018 data sample (denoted Run 2, about 6\invfb) with $\sqrt{s}=13\tev$. The $b$-quark production cross-section in Run 2 is about twice that of Run1~\cite{LHCb-PAPER-2016-031}.

\section{Detector and simulation}
\label{sec:Detector}

The \lhcb detector~\cite{LHCb-DP-2008-001,LHCb-DP-2014-002} is a single-arm forward spectrometer covering the \mbox{pseudo-rapidity} range $2<\eta <5$, designed for the study of particles containing \bquark or \cquark quarks. 
The detector includes a high-precision tracking system consisting of a silicon-strip vertex detector surrounding the $pp$ interaction region, a large-area silicon-strip detector located upstream of a dipole magnet with a bending power of about $4{\mathrm{\, Tm}}$, and silicon-strip detectors and straw drift tubes placed downstream of the magnet. The tracking system provides a measurement of the momentum, \ptot, of charged particles with a relative uncertainty that varies from 0.5\% at low momentum to 1.0\% at 200\gevc. The minimum distance of a track to a primary vertex (PV), the impact parameter (IP),  is measured with a resolution of $(15+29/\pt)\mum$, where \pt is the component of the momentum transverse to the beam, in\,\gevc. Different types of charged hadrons are distinguished using information from two ring-imaging Cherenkov (RICH) detectors. Photons, electrons, and hadrons are identified by a calorimeter system consisting of scintillating-pad and pre-shower detectors, an electromagnetic and a hadronic calorimeter. Muons are identified by a system composed of alternating layers of iron and multiwire proportional chambers.

The online event selection is performed by a trigger system, which consists of a hardware stage, based on information from the calorimeter and muon systems, followed by a software stage, which applies a full event reconstruction. At the hardware trigger stage, events are required to have a muon with high \pt or a hadron, photon, or electron with high transverse energy in the calorimeters. For hadrons, the transverse energy threshold is 3.5\gev. A global hardware trigger decision is ascribed to the reconstructed candidate, the rest of the event or a combination of both; events triggered as such are defined respectively as triggered on signal (TOS), triggered independently of signal (TIS), and triggered on both. The software trigger requires a two-, three- or four-track secondary vertex with a significant displacement from any reconstructed primary vertex. At least one charged particle must have a transverse momentum $\pt > 1.6\gevc$ and be inconsistent with originating from a PV. A multivariate algorithm~\cite{BBDT,LHCb-PROC-2015-018} is used to identify secondary vertices consistent with the decay of a \bquark-hadron.

Simulated events are used to describe the signal and to compute the detection efficiencies. In the simulation, $pp$ collisions are generated using \pythia~\cite{Sjostrand:2007gs} with a specific \lhcb configuration~\cite{LHCb-PROC-2010-056}. Decays of unstable particles are described by the \evtgen package~\cite{Lange:2001uf}, in which final-state radiation is generated using \photos~\cite{Golonka:2005pn}. The interaction of the generated particles with the detector, and its response, are implemented using the \geant toolkit~\cite{Allison:2006ve} as described in Ref.~\cite{LHCb-PROC-2011-006}.

\section{Event selection}
\label{sec:selection}

Candidate $B^0_{(s)}$ signal decays are reconstructed by combining $\Dzb$ candidates, identified via the decay to $\Kp\pim$ tracks, with $\phi\to\Kp\Km$ candidates. The normalisation mode $\Bz\to\Dzb\Kp\Km$ is reconstructed with decays to the same final state.

The selection criteria largely follow those described in Refs.~\cite{LHCb-PAPER-2018-014,LHCb-PAPER-2018-015}, with some improvements to increase the efficiency of the $\Dtphi$ selection. All the charged tracks used in the reconstruction of the $B_{(s)}$ meson candidate are required to be inconsistent with originating from any candidate PVs in the event. The charged kaons and pions are identified using information from the RICH detectors~\cite{Albrecht:2012is,LHCbRICHGroup:2012mgd}. 
To ensure efficient particle identification, kaons and pions are required to have momenta in the range 3 to 100~\gevc, transverse momenta larger than 100\mevc and be within the acceptance of the RICH detectors.
The $\Dzb$ decay products are required to form a good quality vertex with an invariant mass within 25\mevcc of the known $\Dzb$ mass~\cite{PDG2022}. The $\Dzb$ and two kaon candidates from a $\phi$ resonance must also form a good vertex. The reconstructed $\Dzb$ and \PB vertices are required to be significantly separated from any PV candidates. The $B$ invariant-mass resolution is improved by a kinematic fit~\cite{Hulsbergen:2005pu}, where the $\Dzb$ mass is constrained to its known value and the $B$ momentum is constrained to point back to the PV with the smallest value of $\chisqip$, where $\chisqip$ is defined as the difference in the vertex fit $\chisq$ of a given PV reconstructed with and without the particle under consideration. The significance of the distance along the beam direction between the $B$ and $D$ decay vertices, defined as $(z_D-z_B)/\sqrt{\sigma^2_{z_D}+\sigma^2_{z_B}}$, where $z_{D(B)}$ is the position on $z$-axis of the decay
vertex of $\Dz(\Bz)$, and $\sigma_{z_{D(B)}}$ is the corresponding uncertainty,
is used to reject background from charmless decays of $b$-hadrons, such as the $\Bz\to\Kp\pim h^+h^-$ decay ($h$ stands for any hadron). This requirement also suppresses background from charmed mesons directly produced at the PV. The background due to $\Bz\to\theDstarm\Kp, \theDstarm\to\Dzb\pim$ decays is removed by requiring the reconstructed mass difference $m_{\Dzb\pim}-m_{\Dzb}$ to be at least $4.8\mevcc$ away from the charged $\pi$ mass~\cite{PDG2022}, 
when one of the kaons is assumed to be a pion.
The $\Bs\to\Dsm(\to\phi\pi)\Kp$ and swapped background (decay products of $D$ and $\phi$ mesons swapped during reconstruction) are vetoed if the reconstructed $\Dsm$ mass is in the range [1950, 1990]\mevcc and the swapped $\Dzb$(\ie $\Kp\pim$ $\leftrightarrow$  $\pip\Km$) candidates are reconstructed in the range [1840, 1890]\mevcc. Only candidate $B^0_{(s)}\to\Dzb\Kp\Km$ decays with invariant masses in the range [5000, 6000]\mevcc are retained by the selection.

Particle identification (PID) requirements reduce misidentification backgrounds and are optimised with simulation samples and data from the upper sideband region (\mbox{$m_{\Dzb\Kp\Km}\in[5600, 6000]\mevcc$}), where events with one or two $\pi\to K$ misidentifications are localised. This background is low for the signal mode due to the $\phi$ mass requirement.
More stringent PID requirements are imposed on the normalisation mode $\Bz\to\Dzb\Kp\Km$, which is subject to larger contributions from the misidentified background modes $B^0_{(s)}\to\Dzb K\pi$.

A multivariate analysis (MVA) based on a Fisher discriminant~\cite{Fisher:1936et} is used to further separate signal from combinatorial background. The Fisher classifier is trained on simulated $\Bs\to\Dzb\phi$ signal decays and a background-enriched data sample obtained from the upper sideband regions $m_{\Kp\Km}\in[2m_K, \ 1200]\mevcc\cup m_{D\pip\pim}\in[5350, \ 6000]\mevcc$, where $m_K$ is the known kaon mass~\cite{PDG2022} and the $m_{D\pip\pim}$ mass is computed when the two kaon associated tracks are reconstructed using the pion mass hypothesis.  Unlike the $m_{\Dzb\Kp\Km}$ upper sideband, the $m_{D\pip\pim}$ high-mass sideband contains only combinatorial background. The discriminant uses the following information: the smallest values of $\chisqip$ and \pt of the decay products from the $B$-decay vertex; the $\chisq$ probability of the $B$ vertex fit; the $B$ and $\Dzb$ $\chisqip$, and the signed minimum cosine of the angle between the direction of one of the charged tracks from the $B$ decay and the $\Dzb$ meson, as projected in the plane perpendicular to the beam axis. At the MVA selection point corresponding to the largest statistical significance (defined as $\rm{N_{sig}/\sqrt{N_{sig}+N_{bkg}}}$), the signal efficiency is about 95\% and the fraction of rejected background is about 60\%.

After all selection requirements are applied, less than 1\% of the events contain multiple candidates. In these events, a single candidate is retained based on the fit quality of the $B$- and $D$-meson vertices and on the PID information of the $\Dzb$ decay products. The multiple candidate selection is found to have a negligible effect on the final results~\cite{Koppenburg:2017zsh}. Signals for $B^0_{(s)}$ and $\phi$ mesons are observed in the invariant-mass distributions after all of the above selection criteria are applied, as shown in Figs.~\ref{fig:DKKfit} and \ref{fig:phifit}. Finally, $B^0_{(s)}\rightarrow \Dtphi$ candidates are retained if they satisfy $m_{\Kp\Km}\in[2m_K, \ 2m_K+90]\mevcc$.

\section{Fit to the normalisation mode}
\label{sec:DKKfit}

\begin{figure}[t]
  \begin{center}
    \includegraphics[width=0.49\textwidth]{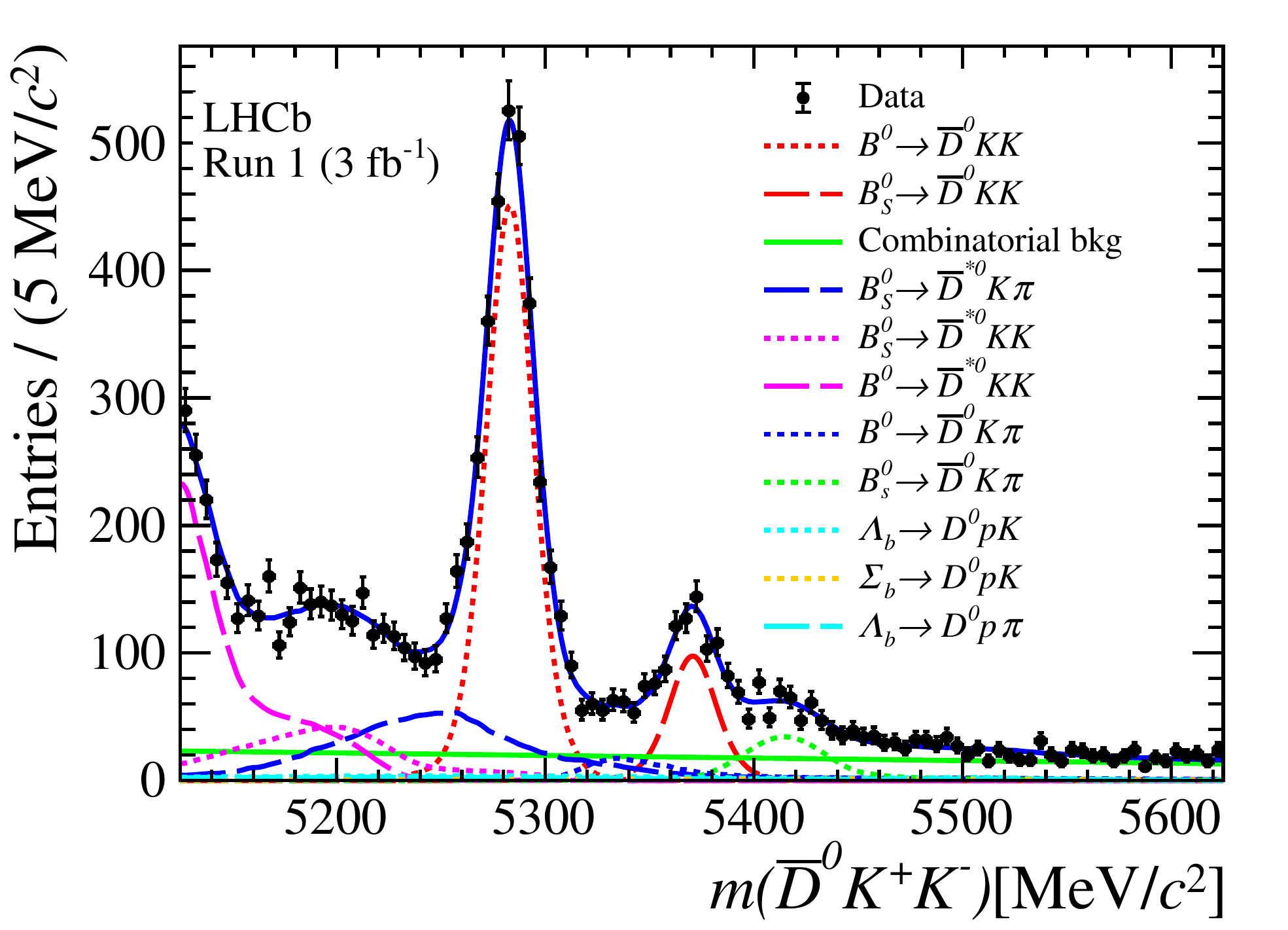}
    \includegraphics[width=0.49\textwidth]{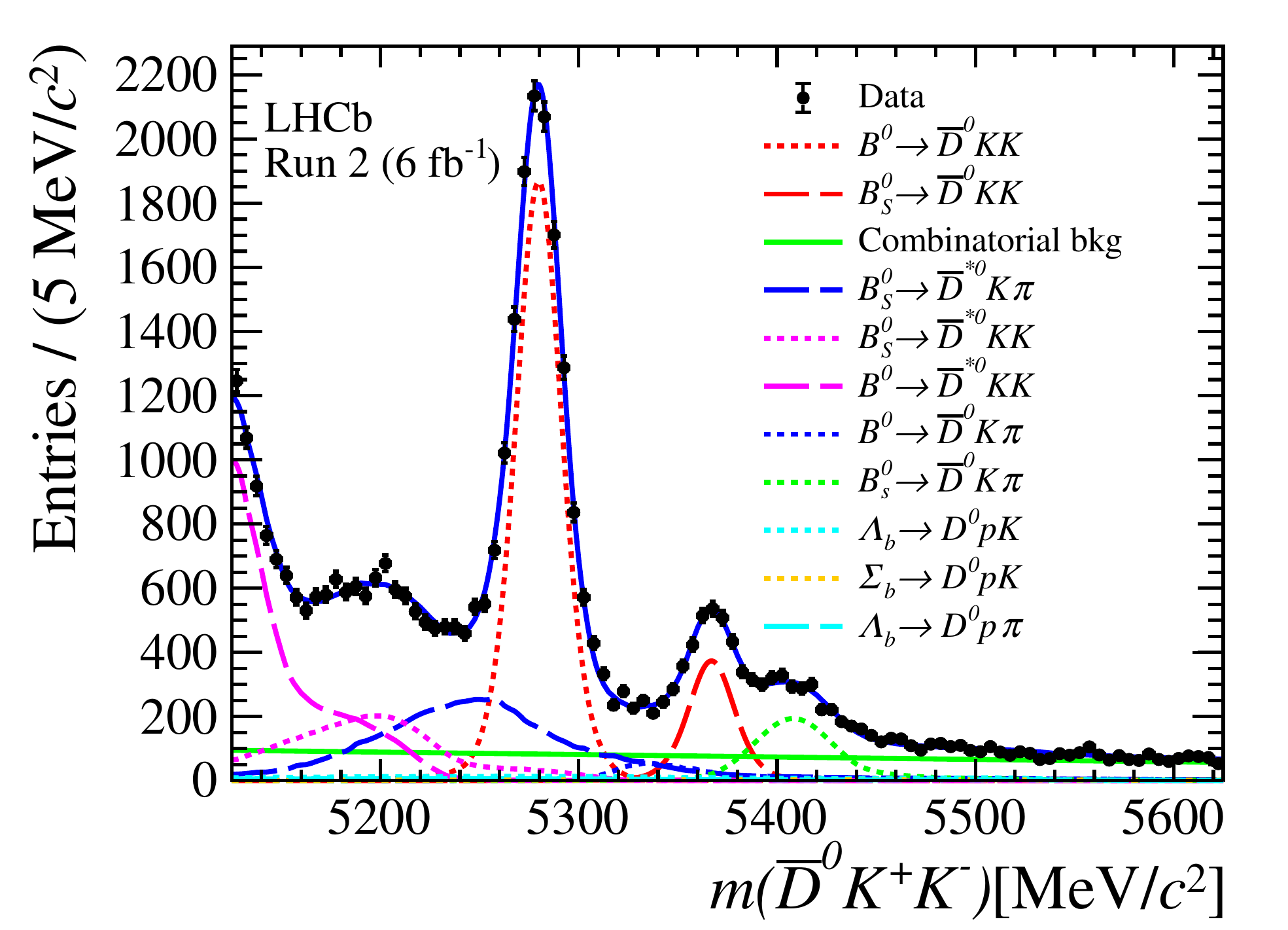}
  \end{center}
  \caption{ Distributions of $m_{\Dzb\Kp\Km}$ for the (left) Run~1 and (right) Run~2 data samples. The results of the fits are overlaid.
   } 
  \label{fig:DKKfit}
\end{figure}

The $\Dzb\Kp\Km$ mass distribution after all selections is shown in Fig.~\ref{fig:DKKfit} and exhibits narrow peaks corresponding to the $\Bz\to\Dzb\Kp\Km$ and $\Bs \to\Dzb\Kp\Km$ decays, as well as several identifiable specific background contributions. Background contributions originate from random particle combinations (combinatorial), particle misidentification, and partially reconstructed $b$-hadron decays.

The lineshape of the reconstructed $\Bz$ mass is parameterised as the sum of two Crystal Ball~(CB)~\cite{Skwarnicki:1986xj} functions with a common mean that is left unconstrained in the fit.
Since the $\Bz$ and $\Bs$ states have similar masses and widths, the same functions are used to describe the $\Bs\to\Dzb\Kp\Km$ shape. The mass difference $\Delta m_B=m_{\Bs}-m_{\Bz}$ is fixed to the world average value~\cite{PDG2022}.
The combinatorial background is modelled with a linear function. 
The shapes of the misidentified and partially reconstructed components~(potentially arising from $\Bz\to\Dzb\Kp\pim$, $\Bs\to\Dzb\Km\pip$, $B^0_{(s)}\to\Dstarzb\Kp\Km$, and $B^0_{(s)}\to\Dstarzb K\pi$ decays) are modelled by non-parametric probability density functions (PDF)~\cite{Cranmer_2001} built from large simulation samples, to which the same selection criteria as for the signal channel are applied. The simulated samples of $\Bz\to\Dzb\Kp\pim$ and $\Bs\to\Dzb\Km\pip$ decays are also assigned weights, depending on the variables $m^2_{\Dzb K^{\pm}}$ and $m^2_{\Dzb\pi^{\mp}}$, using the models and resonances studied in Ref.~\cite{LHCb-PAPER-2014-036,LHCb-PAPER-2015-017}. The normalisation parameters of these background components are left unconstrained during the fit.
Based on the simulation, the $m_{\Dzb\Kp\Km}$  distributions of $\Lb/\Xib\to\Dz p\Km$ and $\Lb\to\Dz p\pim$ backgrounds are broad. 
However, these contributions have prominent peaks in the $\Lb$ mass distribution,
which allows them to be determined precisely from data.
The strategy for modelling these backgrounds closely follows that of the previous $\lhcb$ publication~\cite{LHCb-PAPER-2018-014}: shapes are modelled by non-parametric PDFs from large simulation samples. The estimated yields for those backgrounds are determined from simultaneous fits to the $\Dz p\Km$ and the $\Dz p\pim$ invariant-mass distributions obtained by applying the proton or pion mass hypotheses to one or both kaons of the $\Dzb\Kp\Km$ candidate. Those yields are then used as Gaussian constraints to fit the ${\Dzb\Kp\Km}$ invariant mass.

An extended unbinned maximum-likelihood fit to the $\Dzb\Kp\Km$ invariant-mass distribution, in the range $m_{\Dzb\Kp\Km}\in[5000, \ 6000]\mevcc$, is performed using a global function consisting of the sum of the eleven contributions described above. The fits use the MINUIT~\cite{James:1994vla} minimisation algorithms as implemented in RooFit~\cite{Verkerke:2003ir}. The normalisation channel yields are found to be $2728\pm80$ and $11485\pm170$ for the Run~1 and Run~2 dataset, respectively. The fitted signal yield for Run~1 is 40\% larger than that in the previous analysis~\cite{LHCb-PAPER-2018-014}, owing to optimised selections.

\section{Fit to the signal modes}
\label{sec:Dphifit}

\begin{figure}[tb]
  \begin{center}
    \includegraphics[width=0.49\textwidth]{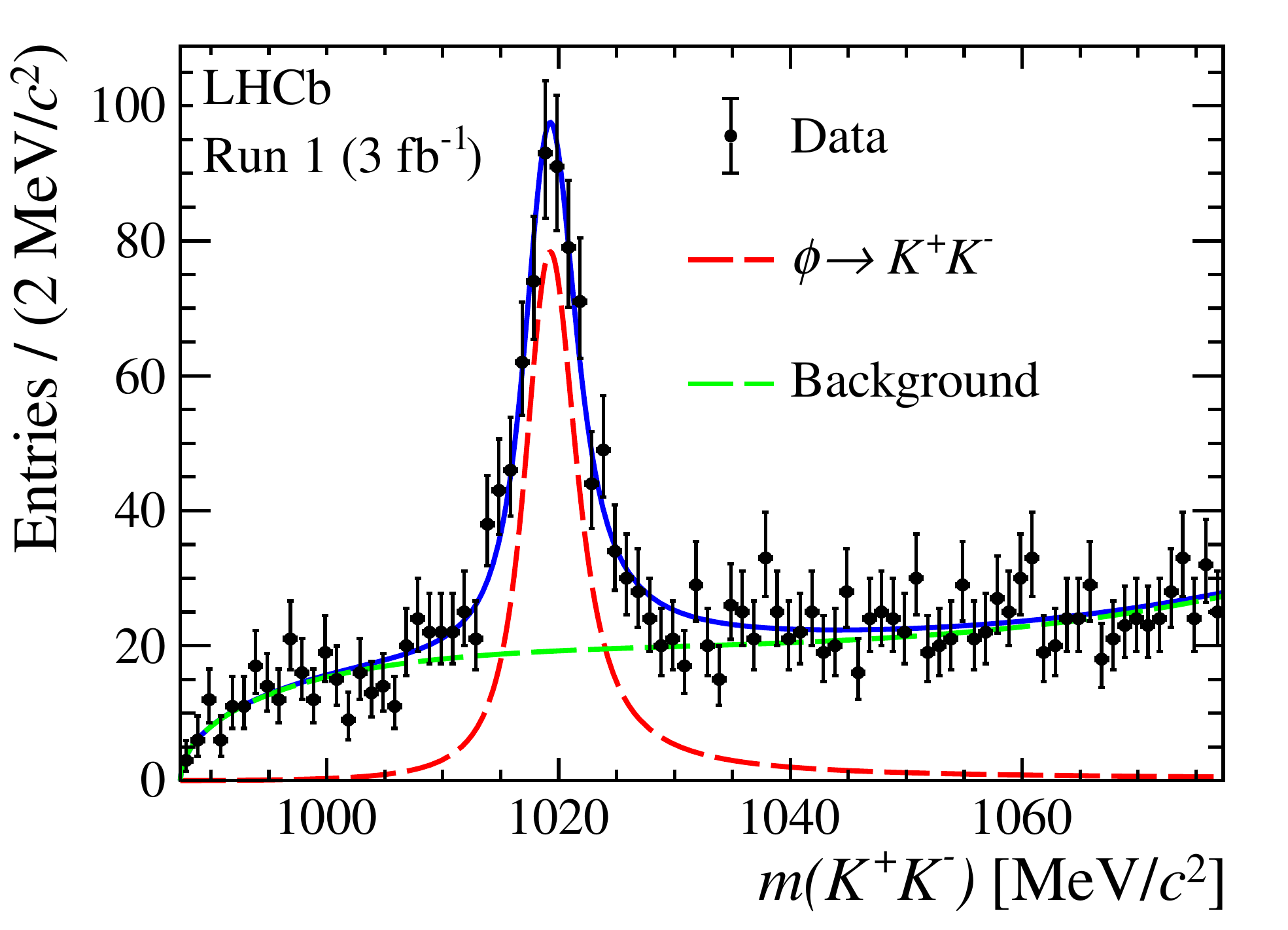}
    \includegraphics[width=0.49\textwidth]{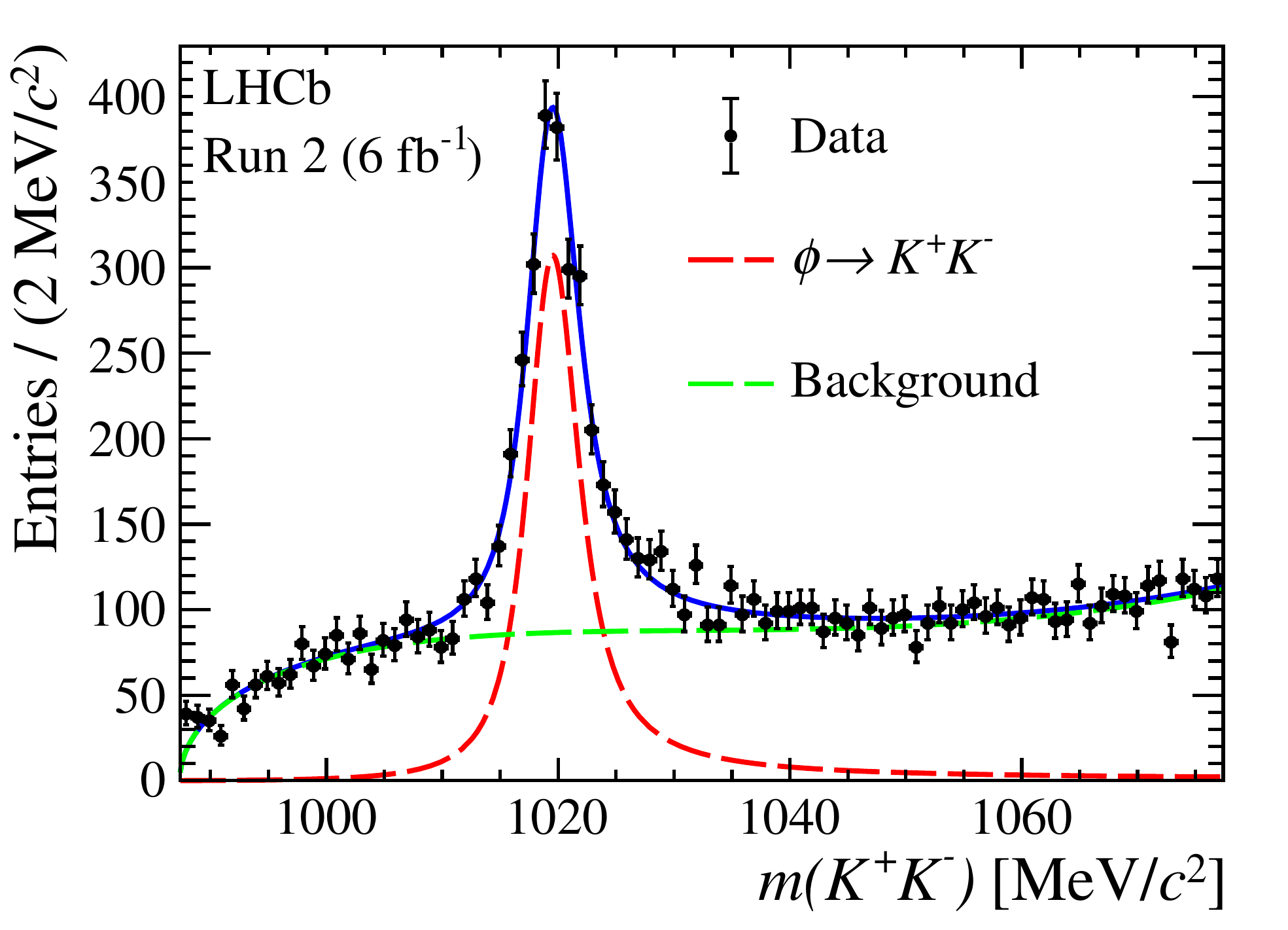}
  \end{center}
  \caption{ Distributions of $m_{\Kp\Km}$ for (left) the Run 1 data and (right) the Run 2 data. The results of the fits are overlaid.
    }
  \label{fig:phifit}
\end{figure}

The method to determine the $B^0_{(s)}\to\Dtphi$ yields proceeds in two sequential steps: fitting the $m_{\Kp\Km}$ distribution and subsequently fitting to the $\Dzb\Kp\Km$ distributions with weights (sWeights). 
The $sPlot$ technique~\cite{Xie:2009rka,Pivk:2004ty} is employed to determine the sWeights from a fit to the $m_{\Kp\Km}$ distribution near the $\phi$ resonance~(\ie $m_{\Kp\Km}\in[2m_K,\ 2m_K+90]\mevcc$~\cite{PDG2022}), that effectively separates $\phi$ signal from non-$\phi$ background (S-wave interference is ignored here according to \cite{lhcb-paper-2012-040}).
The small correlation between the $m_{\Kp\Km}$ and $m_{\Dzb\Kp\Km}$ variables, less than 3\%, ensures that the $sPlot$ technique is appropriate. 
The $\phi$ signal distribution is modelled with a Breit--Wigner PDF convolved with a resolution function described by the sum of two CB functions. The width of the Breit--Wigner function is fixed to the known value~\cite{PDG2022} and the pole mass is determined from data. The parameters of the CB functions are fixed to the values obtained from the simulation. The non-$\phi$ background shape~\cite{LHCb-PAPER-2018-015} is modelled with a phase space factor $p\times q$ multiplied by a quadratic function $1+ax+b(2x^2-1)$, where $p$ and $q$ are the kaon momentum in the $\Kp\Km$ rest frame and the momentum of the $\Dzb$ in the $\Dzb\Kp\Km$ rest frame, respectively. The variable $x$ is defined as $2(m_{\Kp\Km}-2m_{K})/\Delta-1$, where $\Delta$ is the width of the $m_{\Kp\Km}$ mass range considered. The parameters $a$ and $b$ are determined in the fit. The fit results are shown in Fig.~\ref{fig:phifit}. The $\phi$ signal yields obtained from the fit are $615\pm37$, for the Run 1 data, and $2410\pm74$, for the Run 2 data, respectively.

\begin{figure}[t]
  \begin{center}
    \includegraphics[width=0.49\textwidth]{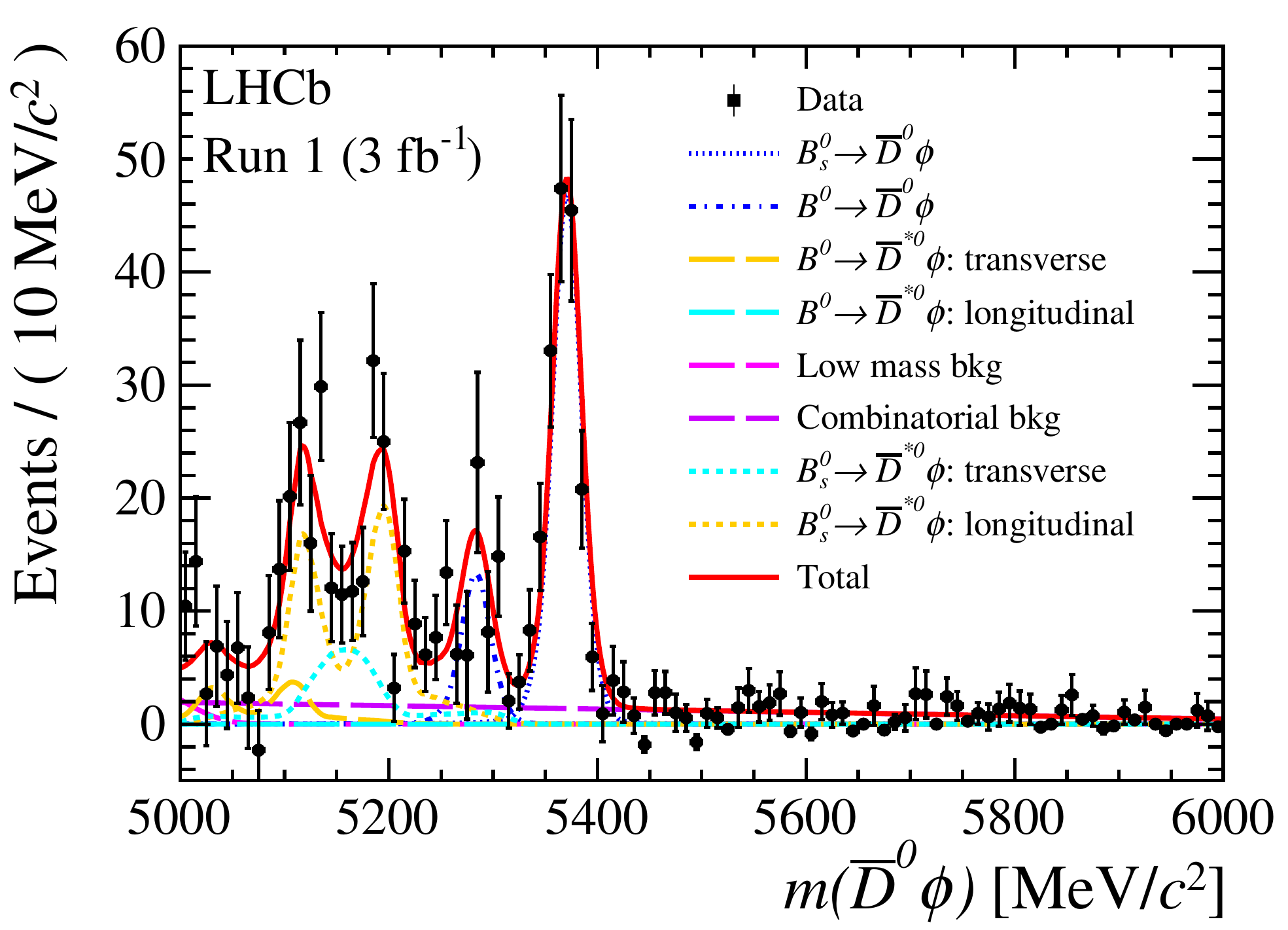}
    \includegraphics[width=0.49\textwidth]{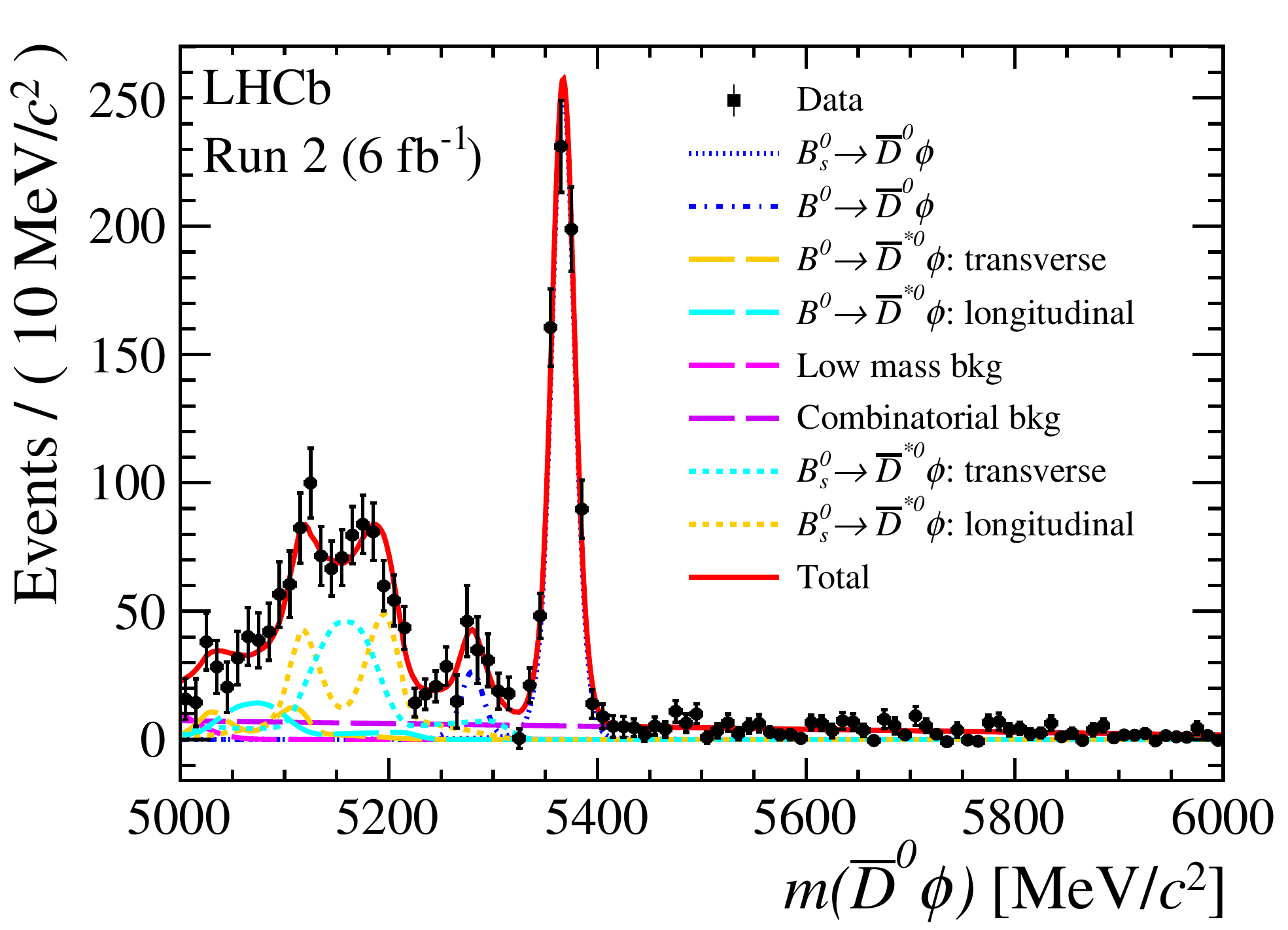}
  \end{center}
  \caption{ Distributions of weighted $m_{D\phi}$ for (left) Run 1 data and (right) Run 2 data. The results of the fits are overlaid. }
  \label{fig:swfit}
\end{figure}

The weighted invariant-mass distributions of $m_{\Dzb\Kp\Km}$ candidates are shown in Fig.~\ref{fig:swfit}. 
A $\Bs\to\Dzb\phi$ signal peak is clearly visible, while the $\Bz\to\Dzb\phi$ signal peak is much less significant. Contributions from $B^0_{(s)}\to\Dstarzb(\to\Dzb\piz/\Dzb\gamma)\phi$ decays can be seen in the region below $m_{\Bs}-m_{\Dstarzb}+m_{\Dzb}$. 

An extended unbinned maximum-likelihood fit to the weighted $m_{\Dzb\Kp\Km}$ distribution is performed to determine the various signal yields. The $\Bs\to\Dzb\phi$ shape is modelled similarly to the normalisation mode $\Bz\to\Dzb\Kp\Km$, for which the mean value and resolution of the PDF are free parameters. The parameters for the $\Bz\to\Dzb\phi$ shape are shared with the $\Bs\to\Dzb\phi$ shape except for the mean value, which is shifted by the known mass difference between the $B_s^0$ and $B^0$ mesons~\cite{PDG2022}. The reconstructed $m(\Dzb\phi)$ mass from the $B^0_{(s)}\to\Dstarzb\phi$ modes strongly depends on the polarisation of the $B$-decay amplitudes, which is a priori unknown. To ascertain its effect, two extreme polarisation configurations are considered: fully longitudinal (the longitudinal polarisation
fraction $f_L=1$) and fully transverse ($f_L=0$). They have different shapes due to the different $\Dstarzb$ helicities, and are modelled by analytical functions derived from their angular distributions~\cite{LHCb-PAPER-2017-021}, with parameters determined from fits to corresponding simulated samples. The PDFs of the two $\Dstarzb$ decay modes $\Dstarzb\to\Dzb\piz/\Dzb\gamma$ are summed 
according to their relative branching fractions~\cite{PDG2022} and corresponding efficiencies. The total PDF for the $\Bs\to\Dstarzb\phi$ signals is then modelled as the sum of the longitudinal and transverse components with $f_L(B_s^0)$ as a free parameter. The $\Bz\to\Dstarzb\phi$ shapes are modelled in a similar way, after applying the known mass shift $m_{\Bs}-m_{\Bz}$ and with an independent parameter $f_L(B^0)$. 
The combinatorial background is accounted for by a linear function. A partially reconstructed background in the low mass region is also considered with its shape determined from the RapidSim~\cite{Cowan:2016tnm} package.

A simultaneous fit to the Run 1 and Run 2 data samples is performed to extract the ratios of the branching fractions between the signal and normalisation modes, defined as 
\begin{align}
\mathcal{R}(B^0_{(s)}\to\Dtphi)&\equiv\frac{\BF(B^0_{(s)}\to\Dtphi)\times \BF(\phi \to \Kp \Km)}{\BF(\BzDbkk)} \nonumber \\
&=\frac{N(B^0_{(s)}\to\Dtphi)\times\varepsilon(\BzDbkk)\times\BF(\phi \to \Kp \Km)}{N(\BzDbkk)\times\varepsilon(B^0_{(s)}\to\Dtphi)}\times F,
\label{DefR}
\end{align}
where $N$ represents the yields of the signal and control channels, and $\varepsilon$ the corresponding efficiencies~(see Sec.~\ref{sec:eff}). The parameter $F$ is unity for the $\Bz\to\Dtphi$ decays, and is $f_d/f_s$ (for Run1, $f_s/f_d=0.239\pm0.008$, and for Run2, $f_s/f_d=0.254\pm0.008$)~\cite{LHCb-PAPER-2020-046} for the $B^0_{s}\to\Dtphi$ decays.  
The results of the simultaneous fit are shown in Table~\ref{tab:simfit}, where the uncertainties are statistical. $f_L(\Bz)$ is not listed due to large statistical fluctuations. More data is needed to determine $f_L(\Bz)$. However, the robustness of these reported results is validated. 

\begin{table}[htbp]
\caption{Results of the simultaneous fit to the $m_{D\phi}$ invariant-mass distribution with $\phi$-sWeighted data. The uncertainties are after corrections using pseudoexperiments~(uncertainties before corrections are shown in parentheses).
}
\normalsize
\begin{center}
\begin{tabular}{lcc}
\hline
Observable  &  Run 1   & Run 2 \\
\hline
\noalign{\vskip 1mm}  
$N(\Bs\to\Dzb\phi)$ & $174.5\pm17.0 ~(13.9)$ 
 &  $740.3\pm34.0~ (28.5)$ \\ 
$N(\Bz\to\Dzb\phi)$ & $48.3\pm19.5$$\,\,(9.3)$ 
 & $\,\,\,77.4\pm26.2~ (13.0)$ \\
$N(\Bs\to\Dstarzb\phi)$ & $233.2\pm25.0~ (20.3)$ 
& $877.5\pm50.0~ (40.8)$ \\
$N(\Bz\to\Dstarzb\phi)$ & $\,\,\,32.4\pm23.3~ (13.9)$ 
& $240.0\pm51.7~ (31.0)$  \\ 
\hline
\noalign{\vskip 1mm}  
$\mathcal{R}(\Bs\to\Dzb\phi)$~(\%) &  \multicolumn{2}{c}{$19.2\pm0.8~(0.7)$} \\
$\mathcal{R}(\Bz\to\Dzb\phi)$~($10^{-3}$)  & \multicolumn{2}{c}{$\,\,\,6.4\pm1.7~(0.9)$} \\
$\mathcal{R}(\Bs\to\Dstarzb\phi)$~(\%)  & \multicolumn{2}{c}{$26.5\pm1.4~(1.1)$} \\
$\mathcal{R}(\Bz\to\Dstarzb\phi)$~(\%)  & \multicolumn{2}{c}{$\,\,\,1.8\pm0.4~(0.2)$} \\
$f_L(\Bs\to\Dstarzb\phi)$~(\%)       & \multicolumn{2}{c}{$53.1\pm6.0~(4.8)$} \\
\hline
\noalign{\vskip 1mm} 
$\BF(\Bs\to\Dzb\phi)$~($10^{-5}$) &  \multicolumn{2}{c}{$2.30\pm0.10$} \\
$\BF(\Bs\to\Dstarzb\phi)$~($10^{-5}$)  & \multicolumn{2}{c}{$3.17\pm0.16$} \\
$\BF(\Bz\to\Dzb\phi)$~($10^{-7}$)  & \multicolumn{2}{c}{$7.7\pm2.1$} \\
$\BF(\Bz\to\Dstarzb\phi)$~($10^{-6}$)  & \multicolumn{2}{c}{$2.2\pm0.5$} \\
\hline
\end{tabular} \end{center}
\label{tab:simfit}
\end{table}

The fit strategy is validated with pseudoexperiments. The mean values of the parameters are confirmed to be unbiased.
However, the uncertainties are underestimated due to the $sPlot$ method used to subtract the non-$\phi$ background. The corrected uncertainties, determined from the pull distributions of the pseudoexperiments, are shown in Table~\ref{tab:simfit}.   

The significance is evaluated with a likelihood-based test, in which the likelihood distribution of the background-only hypothesis is obtained using pseudoexperiments~\cite{Cowan:2010js}, and then convolved with a Gaussian distribution to include the systematic uncertainty. 
Significances of 3.6$\,\sigma$ for the decays $\Bz\to\Dzb\phi$ and 4.3$\,\sigma$ for $\Bz\to\Dstarzb\phi$ are obtained.

\section{Efficiency determination}
\label{sec:eff}

The same strategy is employed to determine the efficiencies as in the previous analysis~\cite{LHCb-PAPER-2018-015}.
The total efficiency 
is the product of the detector
acceptance, reconstruction and selection efficiency, particle identification efficiency and trigger efficiency. The acceptance efficiency corresponds to the fraction of simulated decays reconstructed within the \lhcb detector. The selection efficiency accounts for the software selection in the trigger system,  
the initial selection, the Fisher discriminant selection efficiencies, and the reconstruction of the charged tracks. It is determined with simulated samples except for the track reconstruction part.
The track reconstruction efficiencies are obtained from simulation and corrected using control samples from data. The PID and hardware trigger efficiencies are determined from calibration samples where the abundant $\Dz\to\Km\pip$ sample is used~\cite{LHCb-PUB-2016-021}. 

The simulated samples of the normalisation mode are generated uniformly over the phase-space of the three-body $\Bz\to\Dzb\Kp\Km$ decays and are different from the three-body distributions in the data. 
This effect is taken into account by weighting the simulation samples to match the distributions found in the data and the average efficiency is calculated. 
The total efficiencies of Run~1 and Run~2 for the signal and control channels are shown in Table~\ref{tab:toteff}.

\begin{table}[htbp]
\caption{Summary of total efficiencies for the signal modes $\Bs\to\Dzb\phi$, $\Bs\to\Dstarzb\phi$ and the control mode $\Bz\to\Dzb\Kp\Km$. The efficiencies for $\Bz\to\Dtphi$ decays are assumed to be the same as for the $\Bs\to\Dtphi$ modes. }
\begin{center}
\renewcommand\arraystretch{1.2} 
\begin{tabular}{lcc}
\hline
Mode        & \multicolumn{2}{c}{Efficiency ($\times10^{-4}$)} \\
    & Run~1 & Run~2 \\
\hline
\noalign{\vskip 1mm}  
$\Bz\to\Dzb\Kp\Km$  & $14.55\pm0.08$ & $18.03\pm0.13$ \\
$\Bs\to\Dzb\phi$ & $18.59\pm0.03$ & $23.87\pm0.11$ \\
$\Bs\to\Dstarzb\phi,\Dstarzb\to\Dzb\piz$, longitudinal & $17.08\pm0.13$ & $21.76\pm0.26$ \\
$\Bs\to\Dstarzb\phi,\Dstarzb\to\Dzb\piz$, transverse & $18.01\pm0.12$ & $22.83\pm0.27$ \\
$\Bs\to\Dstarzb\phi,\Dstarzb\to\Dzb\gamma$, longitudinal & $13.61\pm0.11$ & $17.77\pm0.24$ \\
$\Bs\to\Dstarzb\phi,\Dstarzb\to\Dzb\gamma$, transverse  &$14.59\pm0.11$ & $19.10\pm0.25$ \\
\hline
\end{tabular} \end{center}
\label{tab:toteff}
\end{table}

\section{Systematic uncertainties}
\label{sec:systematic}

The same selection criteria are used for all signal decay modes and the normalisation mode, except for the PID requirements that are optimised individually to suppress the different background compositions. Therefore, many potential sources of systematic uncertainty on the efficiencies cancel to a large extent in the ratios of branching fractions displayed in Table~\ref{tab:simfit}. The main sources of systematic uncertainties are due to differences in the trigger and PID efficiencies between the signal and the normalisation channels and in the fit model hypotheses.

The calculation of the hardware trigger efficiency is based on a data-driven method described in Refs.~\cite{RobbeSanchezSChune,LHCb-PAPER-2014-028}, using data from the normalisation channel $\Bz\to\Dzb\Kp\Km$. 
The relative differences in simulation samples between signal and normalisation modes are used to compute the trigger systematic uncertainties, leading to a total relative value of efficiencies equal to 2.4\,\%, for the Run~1 data subset, and 2.3\,\%, for Run~2. Their combined effect gives rise to a relative uncertainty of 2.4\,\% on the branching fractions. The systematic uncertainty on $\varepsilon^{\rm TOS}$ can be neglected since it is determined by the size of the calibration samples, which is much larger than the signal samples. 

The systematic uncertainties associated with the PID requirements are evaluated separately for the different requirements for the $\Kp\Km$ pairs in the $\Bz\to\Dzb\Kp\Km$ and $\Bs\to\Dzb\phi(\Kp\Km)$ decay modes. The uncertainties are attributed to variations of the PID efficiencies by varying the $\pt$ and $\eta$ intervals used for the PID calibration samples. Relative variations of 0.3\,\% for $\Bs\to\Dzb\phi(\Kp\Km)$ and of 2.1\,\% for $\Bz\to\Dzb\Kp\Km$ in efficiencies are observed and assigned as systematic uncertainties. Taking the correlations on the ratios into account, the overall relative PID systematic uncertainty on the ratios $\mathcal{R}(B^0_{(s)}\to\Dtphi)$ is estimated to be 1.8\,\%.

The systematic uncertainties from the signal and background models in the fits are evaluated as follows. 
A number of sources contribute to the signal PDF of the normalisation mode: the modelling of the tails, the different mass resolution of the $\Bz$ and $\Bs$ modes, and the mass difference between $\Bz$ and $\Bs$ mesons, which is fixed in the fit. The values of the tail parameters, estimated from simulated samples, are varied by $\pm\,1\,\sigma$, which gives rise to a 0.3\,\% change in the fitted yield.
An uncertainty of 0.1\% on the event yields is evaluated from an alternative fit in which the mass resolutions of the $\Bz\to\Dzb\Kp\Km$ and $\Bs\to\Dzb\Kp\Km$ modes are assigned to independent parameters, which vary freely, rather than a single shared parameter. Similarly, allowing the mass difference to be unconstrained in the fit induces a relative change of 0.1\,\%. These three sources of systematic uncertainty are considered to be uncorrelated and are added in quadrature to obtain a total systematic uncertainty of 0.3\,\% on the yield of the $\Bz\to\Dzb\Kp\Km$ decay.

The main background components for the control sample are $\Bs\to\Dstarzb\Kp\Km$ and $\ensuremath{\B^0_{(\squark)}}\xspace\to\Dstarzb K\pi$ decays. The $\Bs\to\Dstarzb\Kp\Km$ component is modelled with a non-parametric PDF, based on samples simulated with a square-Dalitz~\cite{LHCB-PAPER-2014-070} plot approach. 
Alternatively, simulated shapes of the main decay modes contributing to the \mbox{$\Bs\to\Dstarzb\Kp\Km$} decay, $\Bs\to D^+_{s1}(2536)\Km$ and $\Bs\to\Dstarzb\phi$, can be used. The difference in the yield obtained with the two fit models is 0.5\%, which is taken as the systematic uncertainty on the $\Bz\to\Dzb\Kp\Km$ yield due to the $\Bs\to\Dstarzb\Kp\Km$ model. Similarly, the component $\ensuremath{\B^0_{(\squark)}}\xspace\to\Dstarzb K\pi$ is modelled with a non-parametric PDF from a square-Dalitz simulated sample of $\ensuremath{\B^0_{(\squark)}}\xspace\to\Dstarzb K\pi$ decays. Alternatively, the shapes of the main decay modes, which are in this case $\Bs\to\Dstarzb\Kstarz$ and $\Bs\to\Ds_1(2536)\pi$, are instead used. The systematic uncertainty on $N(\Bz\to\Dzb\Kp\Km)$ due to the $\Bs\to\Dstarzb K\pi$ model is determined to be 1.5\%. The systematic uncertainties due to background contributions from $\ensuremath{\B^0_{(\squark)}}\xspace\to\Dzb K\pi$ and $\Bz\to\Dstarzb\Kp\Km$ decays are found to be 0.3(0.1)\% and 0.1\%, respectively. A first-order polynomial models the combinatorial background in the baseline fit. To evaluate the systematic uncertainty in the determination of the combinatorial background, as an alternate model, an exponential PDF is used. A relative change on $N(\Bz\to\Dzb\Kp\Km)$ of 0.4\% is found and assigned as the systematic uncertainty.
The background yields from $\Lb$ and $\Xib$ decays have Gaussian constraints in the baseline fit. All sources of systematic uncertainties dealt with above are considered uncorrelated and added in quadrature, which corresponds to a relative systematic error of 1.7\% on $N(\Bz\to\Dzb\Kp\Km)$.

In the fit of the $m_{\Kp\Km}$ distribution, a systematic uncertainty on the $m_{\Kp\Km}$ model is associated with the uncertainty on the $\phi$ mass, which is varied within $\pm\,1\,\sigma$ of its nominal~\cite{PDG2022} uncertainty, and to the background model, that is changed to third order Chebyshev polynomials. In the fit of the weighted $m_{\Dzb\Kp\Km}$ distribution, the systematic uncertainties on the $m_{\Dzb\Kp\Km}$ model are evaluated by using the shape of the normalisation mode $\Bz\to\Dzb\Kp\Km$, described earlier, as an alternate model for the $\Bz\to\Dzb\phi$ and $\Bs\to\Dzb\phi$ decays.
They are also estimated by changing the efficiencies of the decays $B^0_{(s)}\to\Dstarzb\phi$, $\Dstarzb\to\Dzb\piz/\gamma$ with longitudinal or transverse polarisations varied by $\pm\,1\,\sigma$. The differences caused by changing the linear background function to an alternate exponential function are also assigned as systematic uncertainties. Finally, the systematic uncertainty associated with the choice of the relative branching fraction of $\Dstar\to\Dz\piz/\gamma$ decays is calculated by varying its value within its uncertainty~\cite{PDG2022}.
For the estimation of the uncertainty due to the $m_{\Kp\Km}$ model, only the largest variation to the model is used. For the uncertainty due to the $m_{\Dzb\Kp\Km}$ model, they are considered uncorrelated and added in quadrature.

An uncertainty is assigned to the low mass background model, by using the falling tail of a Gaussian distribution for the mass distribution instead of the model from RapidSim.
Uncertainties arising from the size of the simulated signal samples are negligible due to the large size of the generated samples. 
The uncertainty on $f_s/f_d$~\cite{LHCb-PAPER-2020-046}, induces a 3.3\,\% uncertainty on the ratios of branching fractions. 
In addition, as discussed earlier, pseudoexperiments are used to estimate biases on ratios of branching fractions and are accounted for as systematic uncertainties. 
All the systematic uncertainties are summarised in Tables~\ref{tab:totsys} and~\ref{tab:flsys}.

\begin{table}[htbp]
\caption{Summary of systematic uncertainties on the measurements of ratios of branching fractions. 
}
\footnotesize
\begin{center}
\scalebox{0.95}{
\begin{tabular}{lcccc}
\hline 
\noalign{\vskip 1mm} 
Source & $\mathcal{R}(\Bs\to\Dzb\phi)$~(\%) & $\mathcal{R}(\Bz\to\Dzb\phi)$~(\%) & $\mathcal{R}(\Bs\to\Dstarzb\phi)$~(\%) & $\mathcal{R}(\Bz\to\Dstarzb\phi)$~(\%) \\
\hline
Trigger efficiency& 2.4  & 2.4  & 2.4 & 2.4 \\
PID efficiency  & 1.8  & 1.8  & 1.8  & 1.8 \\
$N(\Bz\to\Dzb\Kp\Km)$          & 1.7  & 1.7  & 1.7  & 1.7 \\
Shape model          & 0.6  & 8.1  & 2.2  & 8.1 \\
Low mass background            & 0.1  & 0.9  & 0.8  & 4.8 \\
$\BF(\phi\to\Kp\Km)$           & 1.0  & 1.0  & 1.0  & 1.0 \\
$f_s/f_d$                      & 3.3  & -      & 3.3  & -    \\
Fit bias                       & -  & -  & 0.2  & - \\
\hline 
Total                          & 4.9  & 8.9  & 5.4  & 10.2 \\
\hline

\end{tabular}
}
\end{center}
\label{tab:totsys}
\end{table}

\begin{table}[htbp]
\caption{Summary of systematic uncertainties on the measurements of the fraction of longitudinal polarisation $f_L$ of the $\Bs\to\Dstarzb\phi$ mode.  }
\normalsize
\begin{center}
\begin{tabular}{lc}
\hline 
\noalign{\vskip 1mm} 
Source  & $f_L(\Bs\to\Dstarzb\phi)$~(\%)  \\
\hline
Trigger efficiency          & 0.2   \\
PID efficiency              & 0.1    \\
$N(\Bz\to\Dzb\Kp\Km)$       & -      \\
Shape model                 & 3.4    \\
Low mass background         & 1.2   \\
$\BF(\phi\to\Kp\Km)$        & -      \\
$f_s/f_d$                   & -      \\ 
Fit bias                    & -     \\
\hline
Total                       & 3.6    \\  
\hline
\end{tabular} \end{center}
\label{tab:flsys}
\end{table}

\section{Constraint on the \texorpdfstring{\boldmath$\omega$-$\phi$}{TEXT} mixing angle}
\label{sec:angle}

The $\omega$-$\phi$ mixing angle can be determined by combining the \mbox{$\Bz\to \Dtphi$} and \mbox{$\BzDtomega$~\cite{PDG2022}} branching fractions:
the physical states $\omega$ and $\phi$ can be written as a function of the ideally mixed states $\omega^I\equiv(u\bar{u}+d\bar{d})/\sqrt{2}$ and $\phi^I\equiv s\bar{s}$, as
\begin{equation}
 \begin{pmatrix}
    \omega\\
    \phi
 \end{pmatrix}
    =
 \begin{pmatrix}
    \cos\delta & \sin\delta\\
    -\sin\delta & \cos\delta
 \end{pmatrix}
 \begin{pmatrix}
    \omega^I\\
    \phi^I
 \end{pmatrix}.
\end{equation}
The $\omega$-$\phi$ mixing angle $\delta$ is related to the branching fractions~\cite{Gronau:2008kk} by
\begin{equation}
\tan^2\delta=\frac{\BF(\Bz\to \Dtphi)}{\BF(\BzDtomega)}\times\frac{\Phi(\omega)}{\Phi(\phi)},
\label{tan2delta}
\end{equation}
where $\Phi(\omega)$ and $\Phi(\phi)$ are the integrals of the phase space factors computed over the resonant line shapes. The value of the ratio $\Phi(\omega)/\Phi(\phi)=1.05\pm0.01$~\cite{LHCb-PAPER-2018-015} is used, where the uncertainty comes from the limited knowledge related to the shape parameters of the two resonances.
 
A value of the $\omega$-$\phi$ mixing angle in $\D$ modes is obtained using $\BF(\Bz\to\Dzb\phi)$ (see Sec.~\ref{sec:results}) and the present world average~\cite{PDG2022} $\BF(\Bz\to\Dzb\omega)=(25.4\pm1.6)\times10^{-5}$ 
\begin{equation}
\tan^2\delta_{\D} = (3.2\pm0.9\pm0.3)\times10^{-3},
\end{equation}
where the first uncertainty is statistical,
and the second is systematic. 

The decay $\Bz\to\Dstarzb\phi$ can also be used to calculate the $\omega$-$\phi$ mixing angle. Employing $\BF(\Bz\to\Dstarzb\phi)$ (see Sec.~\ref{sec:results}) and the inputs~\cite{BaBar:2011qjw} $\BF(\Bz\to\Dstarzb\omega)=(45.5\pm4.6)\times10^{-5}$, the following value for the $\omega$-$\phi$ mixing angle for the $\Dstar$ mode is obtained
\begin{equation}
\tan^2\delta_{\Dstar} = (5.0\pm1.1\pm0.5)\times10^{-3},
\end{equation}
where the first uncertainty is statistical,
and the second is systematic. The two decay modes give consistent results at the 1.3\,$\sigma$ level and also agree well with the theoretical predictions~\cite{Gronau:2008kk, Kucukarslan:2006wk, Qian:2008px, Benayoun:2009im, Volkov:2020jor}.

A simultaneous fit is performed to obtain the value of $\delta$ by combining the $\D/\Dstar$ modes, which includes the systematic uncertainties in the branching fraction measurements previously described. The angle is determined to be $\tan^2\delta = (3.6\pm0.7\pm0.4)\times10^{-3}$, where the first uncertainty is statistical and the second is systematic, with a significance of 4.4$\,\sigma$. The value of $|\delta|$ is thus constrained in the range $[3.1,3.8]^{\circ}$ at 68.3\% confidence level (CL) and $[2.1,4.4]^{\circ}$ at 99.7\% CL.

\section{Conclusions}
\label{sec:results}
Studies of $\ensuremath{\B^0_{(\squark)}}\xspace\rightarrow \Dtphi$ decays are performed using the proton-proton collision data collected with the LHCb detector from 2011 to 2018. 
The branching fractions of the $\ensuremath{\B^0_{(\squark)}}\xspace\rightarrow \Dtphi$ decays  are measured relative to the normalisation channel $\Bz\to\Dzb\Kp\Km$. The first evidence for the decays $\Bz\to\Dzb\phi$ and $\Bz\to\Dstarzb\phi$
 is reported with significances of 3.6$\,\sigma$ and 4.3$\,\sigma$, respectively. 
 
Using the signal yields and the efficiencies obtained in previous sections, the branching fraction ratios, defined in Eq.~\ref{DefR}, are measured to be
\begin{align*}
\mathcal{R}(\Bz\to\Dzb\phi)&=(6.4\pm1.7\pm0.6)\times10^{-3},\\
\mathcal{R}(\Bz\to\Dstarzb\phi)&=(1.8\pm0.4\pm0.2)\%,
\end{align*}
where the first uncertainty is statistical, and the second uncertainty is systematic, respectively.  
Using the branching fractions \mbox{$\BF(\Bz\to\Dzb\Kp\Km)=(5.9\pm0.5)\times10^{-5}$} and \mbox{$\BF(\phi\to\Kp\Km)=(49.2\pm0.5)\%$~\cite{PDG2022}}, the measured branching fractions are
\begin{align*}
\BF(\Bz\to\Dzb\phi)=(7.7\pm2.1\pm0.7\pm0.7)\times10^{-7},\\
\BF(\Bz\to\Dstarzb\phi)=(2.2\pm0.5\pm0.2\pm0.2)\times10^{-6}.
\end{align*}
The last uncertainty results from the uncertainties of the branching fractions 
\mbox{$\BF(\Bz\to\Dzb\Kp\Km)$} and $\BF(\phi\to\Kp\Km)$.

 By combining the measured result of the decays $\BzDtomega$, the average $\omega$-$\phi$ mixing angle, $\delta$, is determined to be 
\begin{equation*}
\tan^2\delta = (3.6\pm0.7\pm0.4)\times10^{-3},
\end{equation*}
where the first uncertainty is statistical and the second is
systematic. The value of $|\delta|$ is constrained to be in the range $[3.1,3.8]^{\circ}$ at 68.3\% CL.
This result is consistent with the theoretical prediction in Ref.~\cite{Gronau:2008kk}.

Finally, improved measurements for the branching fractions of the decay modes $\Bs\to\Dtphi$ are obtained
\begin{align*}
\BF(\Bs\to\Dzb\phi)=(2.30\pm0.10\pm0.11\pm0.20)\times10^{-5},\\
\BF(\Bs\to\Dstarzb\phi)=(3.17\pm0.16\pm0.17\pm0.27)\times10^{-5},
\end{align*}
where the first uncertainty is statistical, the second systematic and the third related to the normalisation $\BzDbkk$ mode, including $f_s/f_d$. 
The fraction of longitudinal polarisation of the $\Bs\to\Dstarzb\phi$ decay is computed to be  \mbox{$f_L(\Bs\to\Dstarzb\phi)=(53.1\pm6.0\pm1.9)\%$.} 
These results are consistent with, and supersede, the previous \lhcb measurements~\cite{LHCb-PAPER-2018-015}, which only used the Run~1 data. 
These decay modes, measured here with improved accuracy, can be employed to measure the CKM angle $\gamma$~\cite{Ao:2020cwh}. The measured value of $f_L(\Bs\to\Dstarzb\phi)$ is smaller than the previous measurement~\cite{LHCb-PAPER-2018-015}. However, from Ref~\cite{Ao:2020cwh}, the $\Bs\to\Dstarzb\phi$ modes contribute about 10\%-25\% on the precision of $\gamma$. As a result, the expected $\gamma$ sensitivity is similar to that given in Ref.~\cite{Ao:2020cwh}.


\section*{Acknowledgements}
%
%
\noindent We express our gratitude to our colleagues in the CERN
accelerator departments for the excellent performance of the LHC. We
thank the technical and administrative staff at the LHCb
institutes.
We acknowledge support from CERN and from the national agencies:
CAPES, CNPq, FAPERJ and FINEP (Brazil); 
MOST and NSFC (China); 
CNRS/IN2P3 (France); 
BMBF, DFG and MPG (Germany); 
INFN (Italy); 
NWO (Netherlands); 
MNiSW and NCN (Poland); 
MEN/IFA (Romania); 
MICINN (Spain); 
SNSF and SER (Switzerland); 
NASU (Ukraine); 
STFC (United Kingdom); 
DOE NP and NSF (USA).
We acknowledge the computing resources that are provided by CERN, IN2P3
(France), KIT and DESY (Germany), INFN (Italy), SURF (Netherlands),
PIC (Spain), GridPP (United Kingdom), 
CSCS (Switzerland), IFIN-HH (Romania), CBPF (Brazil),
Polish WLCG  (Poland) and NERSC (USA).
We are indebted to the communities behind the multiple open-source
software packages on which we depend.
Individual groups or members have received support from
ARC and ARDC (Australia);
Minciencias (Colombia);
AvH Foundation (Germany);
EPLANET, Marie Sk\l{}odowska-Curie Actions, ERC and NextGenerationEU (European Union);
A*MIDEX, ANR, IPhU and Labex P2IO, and R\'{e}gion Auvergne-Rh\^{o}ne-Alpes (France);
Key Research Program of Frontier Sciences of CAS, CAS PIFI, CAS CCEPP, 
Fundamental Research Funds for the Central Universities, 
and Sci. \& Tech. Program of Guangzhou (China);
GVA, XuntaGal, GENCAT, Inditex, InTalent and Prog.~Atracci\'on Talento, CM (Spain);
SRC (Sweden);
the Leverhulme Trust, the Royal Society
 and UKRI (United Kingdom).

\addcontentsline{toc}{section}{References}
\ifx\mcitethebibliography\mciteundefinedmacro
\PackageError{LHCb.bst}{mciteplus.sty has not been loaded}
{This bibstyle requires the use of the mciteplus package.}\fi
\providecommand{\href}[2]{#2}

\bibliographystyle{LHCb}
\bibliography{main,standard,LHCb-PAPER,LHCb-CONF,LHCb-DP,LHCb-TDR}

\newpage
\centerline
{\large\bf LHCb collaboration}
\begin
{flushleft}
\small
R.~Aaij$^{32}$\lhcborcid{0000-0003-0533-1952},
A.S.W.~Abdelmotteleb$^{51}$\lhcborcid{0000-0001-7905-0542},
C.~Abellan~Beteta$^{45}$,
F.~Abudin{\'e}n$^{51}$\lhcborcid{0000-0002-6737-3528},
T.~Ackernley$^{55}$\lhcborcid{0000-0002-5951-3498},
B.~Adeva$^{41}$\lhcborcid{0000-0001-9756-3712},
M.~Adinolfi$^{49}$\lhcborcid{0000-0002-1326-1264},
P.~Adlarson$^{77}$\lhcborcid{0000-0001-6280-3851},
H.~Afsharnia$^{9}$,
C.~Agapopoulou$^{43}$\lhcborcid{0000-0002-2368-0147},
C.A.~Aidala$^{78}$\lhcborcid{0000-0001-9540-4988}
% ...

{\footnotesize \it

% ...
$^{9}$Universit{\'e} Clermont Auvergne, CNRS/IN2P3, LPC, Clermont-Ferrand, France\\
% ...
$^{32}$Nikhef National Institute for Subatomic Physics, Amsterdam, Netherlands\\
% ...
$^{41}$Instituto Galego de F{\'\i}sica de Altas Enerx{\'\i}as (IGFAE), Universidade de Santiago de Compostela, Santiago de Compostela, Spain\\
% ...
$^{43}$European Organization for Nuclear Research (CERN), Geneva, Switzerland\\
% ...
$^{45}$Physik-Institut, Universit{\"a}t Z{\"u}rich, Z{\"u}rich, Switzerland\\
% ...
$^{49}$H.H. Wills Physics Laboratory, University of Bristol, Bristol, United Kingdom\\
% ...
$^{51}$Department of Physics, University of Warwick, Coventry, United Kingdom\\
% ...
$^{54}$School of Physics and Astronomy, University of Glasgow, Glasgow, United Kingdom\\
$^{55}$Oliver Lodge Laboratory, University of Liverpool, Liverpool, United Kingdom\\
% ...
$^{63}$Syracuse University, Syracuse, NY, United States\\
% ...
$^{77}$Department of Physics and Astronomy, Uppsala University, Uppsala, Sweden, associated to $^{54}$\\
$^{78}$University of Michigan, Ann Arbor, MI, United States, associated to $^{63}$\\
\bigskip
% ...
}
\end{flushleft}

\end{document}